\input harvmac
\input tables
\def\ra{\rightarrow}
\def\l{\lambda}
\def\ne{\nu_e}
\def\nm{\nu_\mu}
\def\nt{\nu_\tau}
\def\O{{\cal O}}
\def\gsim{{~\raise.15em\hbox{$>$}\kern-.85em
          \lower.35em\hbox{$\sim$}~}}
\def\lsim{{~\raise.15em\hbox{$<$}\kern-.85em
          \lower.35em\hbox{$\sim$}~}} 

\noblackbox
\baselineskip 14pt plus 2pt minus 2pt
\Title{\vbox{\baselineskip12pt
\hbox{hep-ph/9808355}
\hbox{SLAC-PUB-7911}
\hbox{WIS-98/21/Aug-DPP}
}}
{\vbox{
\centerline{Large Mixing and Large Hierarchy Between Neutrinos}
\centerline{with Abelian Flavor Symmetries}
  }}
\centerline{Yuval Grossman\foot{Research supported
by the Department of Energy under contract DE-AC03-76SF00515}}
\medskip
\centerline{\it Stanford Linear Accelerator Center}
\centerline{\it Stanford University, Stanford, CA 94309, USA}
\centerline{yuval@slac.stanford.edu}
\bigskip
\centerline{Yosef Nir and Yael Shadmi}
\medskip
\centerline{\it Department of Particle Physics}
\centerline{\it Weizmann Institute of Science, Rehovot 76100, Israel}
\centerline{ftnir@clever.weizmann.ac.il, yshadmi@wicc.weizmann.ac.il}
\bigskip

\baselineskip 18pt
\noindent
The experimental data on atmospheric and solar neutrinos 
suggest that there is near-maximal mixing between $\nu_\mu$ and
$\nu_\tau$ but that their masses are hierarchically separated. In models
of Abelian horizontal symmetries, mixing of ${\cal O}(1)$ generically
implies that the corresponding masses are of the same order of
magnitude. We describe two new mechanisms by which a large hierarchy
between strongly mixed neutrinos can be achieved in this framework.
First, a {\it discrete} Abelian symmetry can give the desired
result in three ways: mass enhancement,  mixing enhancement and mass 
suppression. Second, {\it holomorphic zeros} can give mass suppression.

\Date{8/98}

\newsec{Introduction}
The Super-Kamiokande collaboration recently announced
evidence for neutrino masses~\ref\SKATM{Y. Fukuda {\it et al.}, 
the Super-Kamiokande Collaboration, hep-ex/9807003}. 
Specifically, various measurements of the flux of atmospheric neutrinos 
can be explained by $\nm\ra\nt$ oscillations with
(for recent analyses, 
see~\nref\GNPW{M.C. Gonzalez-Garcia, H. Nunokawa, O.L.G. Peres and
 J.W.F. Valle, hep-ph/9807305.}%
\nref\BWW{V. Barger, T.J. Weiler and K. Whisnant, hep-ph/9807319.}%
\nref\FLMS{G.L. Fogli, E. Lisi, A. Marrone and G. Scioscia,
 hep-ph/9808205.}%
\refs{\GNPW-\FLMS})
\eqn\ATM{\Delta m_{23}^2\sim2\times10^{-3}\ eV^2\ ,
\ \ \ \sin^22\theta_{23}\sim1\ .}
The solar neutrino flux has been measured by various experiments.
The data from the chlorine, GALLEX, SAGE and
Super-Kamiokande experiments can be explained by $\ne\ra\nu_x$ 
($x=2$ or 3) oscillations with one of the following three options
(for a recent analysis, see 
\ref\BKS{J.N. Bahcall, P.I. Krastev and
 A. Yu. Smirnov, hep-ph/9807216.}):
\eqn\SOL{\matrix{&\Delta m_{1x}^2\ [eV^2]&\sin^22\theta_{1x}\cr
{\rm MSW(SMA)}&5\times10^{-6}&6\times10^{-3}\cr
{\rm MSW(LMA)}&2\times10^{-5}&0.8\cr
{\rm VO}&8\times10^{-11}&0.8\cr}}
Here MSW refers to matter-enhanced oscillations, VO refers to vacuum
oscillations, and SMA (LMA) stand for small (large) mixing angle.
Only central values are quoted for the various parameters.
(We assume that there are no light sterile neutrinos. Otherwise
a light $\nu_s$ could replace $\nt$ in \ATM\ or $\nu_x$ in \SOL.) 
 
With three light massive neutrinos, there are nine new flavor 
parameters in addition to the thirteen of the Standard Model: 
three neutrino masses, three mixing angles and three CP-violating
phases. If the discrepancy between experiments and theory for
both atmospheric and solar neutrinos is indeed explained by
\ATM\ and \SOL, then four of these new parameters have been measured. 
This information has strong impact on models that explain the
observed smallness and hierarchy in flavor parameters.
In this work, we focus on models of Abelian horizontal symmetries. 

A crucial point concerning the combination of \ATM\ and \SOL\
is that the two mass-squared differences are widely separated:
\eqn\AtmSol{10^{-7}\lsim{\Delta m^2_{1x}\over\Delta m^2_{23}}
\lsim10^{-2}\ .}
As a consequence of \AtmSol, there are only two different forms of 
neutrino mass matrices in the charged lepton mass basis
that are consistent with both \ATM\ and \SOL\ 
\ref\BHSSW{R. Barbieri, L.J. Hall, D. Smith, A. Strumia and
 N. Weiner, hep-ph/9807235.}.
One of these forms describes a situation where the parameters relevant 
to $\nm-\nt$ oscillations effectively reside in the $2\times2$ 
sub-matrix of the full $3\times3$ light neutrino mass matrix. 
The $\nm-\nt$ block is then of the form (we ignore CP violation):
\eqn\Mtwo{\eqalign{M^{(2)}=&\ {v^2\over M}\pmatrix{C&B\cr B&A\cr}\ ,\cr
A,B,C=&\ \O(1),\ \ \ |AC-B^2|\ll B^2\ .\cr}}
In addition, there is one form of the $3\times3$ matrix that gives
large $s_{23}\equiv\sin\theta_{23}$ and cannot be effectively reduced 
to the $2\times2$ description of atmospheric neutrinos, that is \BHSSW\
\eqn\Mthree{\eqalign{
M^{(3)}=&\ {v^2\over M}\pmatrix{0&B&A\cr B&0&0\cr A&0&0\cr},\cr
A,B=&\ \O(1).\cr}} 

In models of (supersymmetric) Abelian horizontal symmetries, 
large  $\nm-\nt$ mixing is generically achieved with 
either~\ref\GrNi{Y. Grossman and Y. Nir,
 Nucl. Phys. B448 (1995) 30, hep-ph/9502418.}\
\eqn\MtwoGN{\eqalign{M^{(2)}=&\ {v^2\over M}\pmatrix{C&B\cr B&A\cr},\cr
A,B,C=&\ \ \O(1),\ \ \ AC-B^2=\O(1)\ ,\cr}}
corresponding to equal horizontal charges for the two
lepton doublets, or 
with~\ref\BLPR{P. Binetruy, S. Lavignac, S. Petcov and P. Ramond,
 Nucl. Phys. B496 (1997) 3, hep-ph/9610481.}\
\eqn\MtwoBR{\eqalign{M^{(2)}=&\ {v^2\over M}\pmatrix{c&B\cr B&a\cr}\ ,\cr
B=&\ \O(1),\ \ \ a,c\ll1\ ,\cr}}
corresponding to opposite horizontal charges for the two
lepton doublets. Thus, it is non-trivial to have in this framework
large mixing between neutrinos whose masses are widely separated. 
It is the purpose of this work to describe the various ways by which 
mass matrices of the forms \Mtwo\ or \Mthree\ can arise in models 
of Abelian horizontal symmetries and, in particular, to suggest two new 
methods for generating \Mtwo\ or \Mthree.

The structure of this paper is as follows. In section~2, we clarify some
subtleties concerning the low-energy effective theory for neutrinos
with an approximate flavor symmetry. In section~3, we review previously
proposed mechanisms for inducing a large hierarchy.
We then suggest two new mechanisms for producing a large hierarchy
while maintaining large mixing. One, employing discrete symmetries, 
is described in  section~4. The other, based on holomorphic zeros, 
is discussed in  section~5.
Finally, we present our conclusions in section~6.

\newsec{Neutrino Mass Matrices with Abelian Horizontal Symmetries}
We study Supersymmetric models with an Abelian horizontal symmetry $H$. 
For a simple $H$, we assume that there is a single small breaking parameter. 
Without loss of generality, we assume that the breaking parameter carries 
charge $-1$ under the horizontal symmetry. (By this we mean that the symmetry
is broken spontaneously by a VEV of a scalar, Standard Model singlet field
to which we attribute horizontal charge $-1$. The small parameter is then
the ratio between this VEV and the scale where the information about the
breaking is communicated to the observable 
sector~\ref\FrNi{C.D. Froggatt and H.B. Nielsen, Nucl. Phys. 
B147 (1979) 277.}.)
If the horizontal symmetry is semi-simple, then we assume that each
simple subgroup is broken by a single small parameter.
Well below the $H$-breaking scale $\Lambda_H$, we have an effective theory
with the following selection rules:
\item{(a)} Terms in the superpotential that carry charge $n\geq0$
under $H$ are suppressed by $\O(\lambda^{n})$, while those with $n<0$ 
are forbidden (due to the holomorphy of the 
superpotential~\ref\LNS{M. Leurer, Y. Nir and N. Seiberg,
Nucl. Phys. B398 (1993) 319, hep-ph/9212278.}).
\item{(b)} Terms in the K\"ahler potential that carry charge $n$
under $H$ are suppressed by $\O(\lambda^{|n|})$. 

\nref\RaSi{A. Rasin and J.P. Silva, Phys. Rev. D49 (1994) 20, 
 hep-ph/9309240.}%
In the case of neutrinos, there is however a subtle point concerning
the natural mass scale of the singlet neutrinos. 
The mass of these singlet neutrinos could in principle be lower
than the $H$-breaking scale.
For example, total lepton number may be a symmetry of the full
theory that is broken only at a scale $\Lambda_L$ that is much lower
than the $H$-breaking scale,  $\Lambda_L\ll\Lambda_H$. Then,
the low-energy effective theory below $\Lambda_H$ includes both the
doublet (`left-handed') and singlet (`right-handed') neutrino fields.
We will denote lepton-doublet fields by $L_i$ and singlet neutrino
fields by $N_i$, with $i$ a flavor index. One can further integrate out 
the singlet neutrinos to obtain an effective theory for the left-handed 
neutrinos only, which is valid well below $\Lambda_L$.
The selection rules cannot, however, be simply applied to this effective 
theory. In particular, terms in the superpotential that carry charge 
$n<0$ under $H$ either vanish or are {\it enhanced} by 
$\O(\lambda^{-n})$. The latter possibility arises because the light 
neutrino mass matrix is given by
\eqn\Mlight{M_\nu^{\rm light}=M_D(M_N)^{-1}M_D^T\ ,}
where $M_D$ is the Dirac mass matrix that couples the doublet and singlet
neutrinos, while $M_N$ is the Majorana mass matrix for the singlet neutrinos.
Then $M_\nu^{\rm light}$ depends on negative powers of the heavy neutrino
masses which are themselves suppressed by powers of $\lambda$.
Therefore, in this type of models, one has in general to analyze the
full neutrino mass matrix in order to estimate the flavor parameters
in the light neutrino sector.\foot{If all neutrino charges
are positive, so that there are no holomorphic zeros in $M_D$ and
in $M_N$, one can estimate the order of magnitude of the various
light neutrino masses and mixings independently of the singlet neutrino
charges~\refs{\RaSi,\GrNi}.}

Another possibility is that the scales $\Lambda_H$ and $\Lambda_L$
are the same. This will be the case if there is no
lepton number symmetry in the full theory and if all Standard Model
singlet neutrinos appear in vector representations of $H$. Then
one can study the effective low energy theory with doublet neutrinos
only, and apply the selection rules (a) and (b) given above. In particular,
superpotential terms with negative $H$-charge should be set to zero.
One can always find a full high energy theory (of the type described here,
namely with no singlet fermion representations that are chiral under $H$)
which yields the deduced flavor structure of the light neutrinos.

In the models we present below, whenever we analyze the mass matrix
for both doublet and singlet neutrinos, we implicitly assume that
the full high energy theory is of the first type described above
($\Lambda_L\ll\Lambda_H$), while when we analyze the effective
theory for the doublet neutrinos only, we assume a full high energy
theory of the second type ($\Lambda_L=\Lambda_H$). 

\newsec{Previously Proposed Mechanisms}

\subsec{Enhanced Masses}
In models with $\Lambda_L\ll\Lambda_H$, a neutrino mass could be
enhanced beyond the naive expectation by the see-saw mechanism.
To understand the mechanism in more detail, suppose that one of 
the entries in $M_\nu^{\rm light}$, the effective $3\times3$ matrix for the
light (left-handed) neutrinos, carries a certain horizontal charge $n$.
Assume further that there exists a breaking parameter $\varepsilon$
with charge $m$ such that $p=n/m$ is a {\it positive integer}. 
(The $\varepsilon$ parameter could be either the only breaking parameter 
or one of several breaking parameters. Some or all of the other 
breaking parameters could have a sign opposite to that of $n$.)
In the usual supersymmetric Froggatt-Nielsen mechanism for charged 
fermions, the entry in question cannot depend on $\varepsilon$
because of holomorphy. For neutrinos, however, this entry could
get an {\it enhanced} contribution proportional to $\varepsilon^{-p}$.
Whether this happens depends on whether there are singlet neutrinos 
with masses suppressed by at least $\varepsilon^p$. This idea is 
presented and demonstrated by an explicit example in ref. \BHSSW.

We present a somewhat different model mainly for pedagogical reasons: 
it enables us to elucidate the crucial points 
of this mechanism and the subtleties discussed in section 2. 
Our model, however, cannot be made consistent
with the detailed requirements of~\SOL.

The horizontal symmetry is $U(1)_H$  and it is broken by a small
parameter $\lambda$ to which we attribute charge $-1$. The neutrino 
fields carry the following $H$-charges:
\eqn\LNcharges{L_1(y),\ \ L_2(-x),\ \ L_3(-x),\ \ N_1(a),\ \ N_2(b),\ \ 
N_3(c)\ ,}
with
\eqn\orderLN{y\geq x>0,\ \ \ \ a\geq b\geq c\geq0\ .}
All entries in the $2-3$ block of $M_\nu^{\rm light}$ carry charge
$-2x$. Therefore, each of them could be either zero or enhanced by
$\l^{-2x}$. Which of the two options is realized depends sensitively
on the $N_i$-content of the model. Applying the selection rules to the
$M_D$ and $M_N$ blocks of the full $6\times6$ mass matrix, we get:\foot{
We remind the reader that we only write the dependence of the various 
entries on the small breaking parameters and omit the unknown $\O(1)$ 
coefficients. The latter are arbitrary except that $M_{\nu}^{\rm light}$
and $M_N$ are symmetric.}
\eqn\Mdirac{M_D\sim\vev{\phi_u}\pmatrix{\l^{a+y}&\l^{b+y}&\l^{c+y}\cr
(\l^{a-x})&(\l^{b-x})&(\l^{c-x})\cr (\l^{a-x})&(\l^{b-x})&(\l^{c-x})\cr},}
where the terms in parenthesis are non-zero only for non-negative exponents,
and
\eqn\Mmajor{M_N\sim\Lambda_L\pmatrix{\l^{2a}&\l^{a+b}&\l^{a+c}\cr
\l^{a+b}&\l^{2b}&\l^{b+c}\cr \l^{a+c}&\l^{b+c}&\l^{2c}\cr}.}
One can now distinguish between three interesting cases:
\item{(i)} $a<x$: Only one of the light neutrinos is massive with
mass of order ${\vev{\phi_u}^2\over\Lambda_L}\l^{2y}$, 
while the other two neutrinos are massless. Indeed, in this case
none of the $N_i$ masses is suppressed by as much as $\l^{2x}$, 
so that the holomorphic zeros are maintained. 
\item{(ii)} $x\leq b$: The three light neutrinos are massive with
masses (relative to the scale ${\vev{\phi_u}^2\over\Lambda_L}$) 
of order $\{\l^{-2x},\l^{-2x},\l^{2y}\}$. In this case two 
(or all three) of the $N_i$ have their masses below $\Lambda_L\l^{2x}$, 
thus allowing an enhancement of the two relevant
light neutrino masses.
\item{(iii)} $b<x\leq a$: Two of the light neutrinos are massive with
masses of order $\{\l^{-2x},\l^{2y}\}$ and one neutrino is massless.
Now only one of the $N_i$ has its mass below $\Lambda_L\l^{2x}$, 
and consequently the holomorphic zeros are lifted in such a way that 
only one of the light neutrinos has its mass enhanced by $\l^{-2x}$.

In all three cases, the $\nm-\nt$ mixing is large, namely $s_{23}\sim1$.
Actually, while the $s_{12}^\nu$ and $s_{13}^\nu$ angles in the 
diagonalizing matrix of $M_\nu^{\rm light}$ may be different from 
one case to another, we generically expect those of the charged
lepton sector to be the same in all cases, and consequently we get
for the mixing angles:
\eqn\mixEM{s_{12}\sim\l^{x+y},\ \ \ s_{13}\sim\l^{x+y},\ \ \ 
s_{23}\sim1\ .}

Case (iii) above is  of special interest to us, since together
with the large $s_{23}$ it gives a strong hierarchy between the
corresponding masses.  This case therefore constitutes an example 
of the proposed mechanism: the mass of one combination of neutrinos with 
$H$-charges $-x$ is enhanced by the horizontal
symmetry, while the other is not. 

As mentioned above, the model of case (iii), while pedagogically
useful, cannot satisfy all our phenomenological requirements.
The light neutrino mass matrix, in the basis where the $2-3$ block is 
diagonal, is given by
\eqn\MLEM{M_\nu^{\rm light}\sim{\vev{\phi_u^2}\over\Lambda_L}\pmatrix{
\l^{2y}&0&\l^{x+y}\cr 0&0&0\cr \l^{x+y}&0&\l^{-2x}\cr}\ .}
This structure cannot be made consistent with the VO or the MSW(LMA)
solutions since the mixing angle is, at most, of order 
$(\Delta m^2_{1x}/\Delta m^2_{23})^{1/2} \ll 1$. It is also inconsistent 
with the MSW(SMA) solution since $\nu_e$ is heavier than $\nu_x$ 
(the light $\nu_\mu-\nu_\tau$ combination). With a different
choice of charges, {\it e.g.} the one made in \BHSSW\ ($x=-a=b=c$),
one can get a matrix that is consistent with either of the
large-angle solutions.

\subsec{Neutrino Masses from Different Sources} 
Different neutrino masses could come from different sources, so
that the hierarchy is determined not only by the horizontal charges.
A framework where this is the case is that of supersymmetry without
$R$-parity. The Abelian horizontal symmetry could replace
$R$-parity in suppressing the dangerous lepton-number violating
couplings~\nref\Bank{T. Banks, Y. Grossman, E. Nardi and Y. Nir,
 Phys. Rev. D52 (1995) 5319, hep-ph/9505248.}%
\nref\BGNN{F.M. Borzumati, Y. Grossman, E. Nardi and Y. Nir,
 Phys. Lett. B384 (1996) 123, hep-ph/9606251.}%
\nref\BILR{P. Binetruy, N. Irges, S. Lavignac and P. Ramond,
 Phys.Lett. B403 (1997) 38, hep-ph/9612442.}%
\nref\ChLu{E.J. Chun and A. Lukas,
 Phys.Lett. B387 (1996) 99, hep-ph/9605377.}%
\nref\CCK{K. Choi, E.J. Chun and H. Kim,
 Phys. Lett. B394 (1997) 89, hep-ph/9611293.}%
\nref\Nard{E. Nardi, Phys. Rev. D55 (1997) 5772, hep-ph/9610540.}%
\nref\DPTV{M. Drees, S. Pakvasa, X. Tata and T. ter Veldhuis,
 Phys. Rev. D57 (1998) 5335, hep-ph/9712392.}%
\nref\ElLR{J. Ellis, S. Lola and G.G. Ross, hep-ph/9803308.}%
\refs{\Bank-\ElLR}. If the $B$- and $\mu$-terms are not aligned, so
that a single neutrino acquires a mass from mixing with neutralinos
while the other two acquire masses at the loop level only, then the
hierarchy between the heavier neutrino and the two light ones
is too strong in general to accommodate both \ATM\ and \SOL. 
But if $B$ and $\mu$
are aligned, all three neutrinos get loop-level masses. In a large class 
of such models, large mixing ($s_{23}\sim1$) predicts a hierarchy 
$m_{\nu_2}/m_{\nu_3}\sim m_\tau^4/3m_b^4$.
Recent studies of this framework were made in refs.
\nref\CKKL{E.J. Chun, S.K. Kang, C.W. Kim and U.W. Lim, hep-ph/9807327.}%
\nref\BFK{V. Bednyakov, A. Faessler and S. Kovalenko, hep-ph/9808224.}%
\nref\MRV{B. Mukhopadhyaya, S. Roy and F. Vissani, hep-ph/9808265.}%
\nref\Kong{O.C.W. Kong, hep-ph/9808304.}%
\refs{\CKKL-\Kong}.

Another scenario in which different light neutrinos get their masses
from different sources involves a single singlet
neutrino. Then only one light neutrino acquires its mass at tree level,
while masses for the other two are generated  radiatively.
This possibility was discussed in the context of the recent
Super-Kamiokande data in 
ref.~\ref\DaKi{S.F. King, hep-ph/9806440;
 S. Davidson and S.F. King, hep-ph/9808296.}.

\subsec{Accidental Hierarchy}
It may be that, as far as the small breaking parameters are
concerned, it is model \MtwoGN\ which is realized in nature with  
the parameters of $O(1)$ accidentally giving a small determinant
\nref\BLR{P. Binetruy,  S. Lavignac and P. Ramond,
 Nucl. Phys. B477 (1996) 353, hep-ph/9802334.}%
\nref\ILR{N. Irges,  S. Lavignac and P. Ramond,
 Phys. Rev. D58 (1998) 035003, hep-ph/9802334.}%
\nref\EIR{J.K. Elwood, N. Irges and P. Ramond, hep-ph/9807228.}%
\nref\ELLN{J. Ellis, G.K. Leontaris, S. Lola and D.V. Nanopoulos,
 hep-ph/9808251.}%
\refs{\BLR-\ELLN}.
If this is the case, then we are misled to think that \AtmSol\
is related to an approximate horizontal symmetry. Indeed, the
fundamental parameters are the neutrino masses and not the
masses-squared. The ratio between the masses need only be $\O(0.1)$. 
This is not a very small number
and so it is not impossible that it is accidentally (rather than
parametrically) small. Rather plausible explicit examples
for such a situation were given in refs.~\refs{\ILR,\ELLN}. 
 
\newsec{Discrete Symmetries}
In all explicit examples that we construct in this 
(and in the next) section, we aim at the following order of 
magnitude estimates for the charged lepton masses:
\eqn\chaobs{m_e/m_\mu\sim\l^3,\ \ \ m_\mu/m_\tau\sim\l^2,\ \ \ 
m_\tau/\vev{\phi_d}\sim\l^3.}
The last relation is appropriate for $\tan\beta\sim1$ but all the
models can be easily modified for larger values of $\tan\beta$.
For the neutrinos, we require
\eqn\neuobs{s_{23}\sim1,\ \ \ s_{13}\lsim\l\ ,}
where the bound on $s_{13}$ comes from the combination of CHOOZ and
Super-Kamiokande data~\refs{\BWW,\FLMS}, and
\eqn\SOLobs{\matrix{&\Delta m_{1x}^2/\Delta m_{23}^2&s_{12}\cr
{\rm MSW(SMA)}&\l^3-\l^4&\l^2\cr
{\rm MSW(LMA)}&\l^2-\l^4&1\cr
{\rm VO}&\l^{10}-\l^{12}&1\cr}}

In our models of discrete symmetries, we take
\eqn\Hdis{H=Z_m\times Z_n}
and breaking parameters
\eqn\epsdis{\varepsilon_m=\O(\l),\ \ \ \varepsilon_n=\O(\l)\ .}
The precise value of $n$ is not important except that 
it is large enough so that, for the fermion masses, the symmetry is
effectively $Z_m\times U(1)$. On the other hand, the $Z_m$ symmetry 
is required to allow a mass hierarchy $m_2/m_3\sim\l^m$ 
with $s_{23}\sim1$. We will use then $m=2$ for MSW solutions
and $m=5$ or 6 for VO models.

\subsec{Mass Enhancement}
A discrete symmetry can enhance one of the neutrino masses.
Take, for example, an even $m$ with the following 
horizontal charges of the second and third lepton generations:
\eqn\discha{L_2(m/2-1,1),\ \ \ L_3(m/2,0),\ \ \ \bar\ell_2(m/2+1,4),\ \ \ 
\bar\ell_3(m/2+1,2).}
Then, the $2\times2$ mass matrices have the form
\eqn\dismas{M_\nu^{(2)}\sim{\vev{\phi_u}^2\over\Lambda_H}\pmatrix{
\l^m&\l^m\cr\l^m&1\cr},\ \ \ M_{\ell^\pm}^{(2)}\sim\vev{\phi_d}
\pmatrix{\l^5&\l^3\cr \l^5&\l^3}.}
When we rotate to the charged lepton mass basis, $M_\nu^{(2)}$
assumes the form \Mtwo. In particular, we get
\eqn\dispar{{m_{\nu_2}\over m_{\nu_3}}\sim\l^m,\ \ \ s_{23}\sim1\ .}

The crucial point to notice in \dismas\ is that $(M_\nu^{(2)})_{33}$, which
would have been $\O(\l^m)$ under a $U(1)$, is enhanced to $\O(1)$
under the discrete symmetry.
The idea is then that horizontal charges that would lead to masses of
the same order of magnitude with $H=U(1)$, can lead to hierarchical 
masses if in some of the mass terms the discrete nature of $H=Z_m$ 
comes into play.
A large mixing angle could arise with either symmetry.

To give an example of the MSW(SMA) option, we take $m=2$ and 
the following charges:
\eqn\diMEms{L_1(0,3),\ \ L_2(0,1),\ \ L_3(1,0),\ \ 
\bar\ell_1(0,5),\ \ \bar\ell_2(0,4),\ \ \bar\ell_3(0,2)\ .}
Then, the mass matrices have the form
\eqn\MEmas{M_\nu\sim{\vev{\phi_u}^2\over\Lambda_H}\pmatrix{
\l^6&\l^4&\l^4\cr \l^4&\l^2&\l^2\cr \l^4&\l^2&1\cr},\ \ \ 
M_{\ell^\pm}\sim\vev{\phi_d}
\pmatrix{\l^8&\l^7&\l^5\cr \l^6&\l^5&\l^3\cr \l^6&\l^5&\l^3},}
yielding
\eqn\dismsw{{\Delta m^2_{1x}\over\Delta m^2_{23}}\sim\l^4,
\ \ \ s_{12}\sim\l^2,\ \ \ s_{23}\sim1,\ \ \ s_{13}\sim\l^2.}

A viable VO model is produced by $m=6$ and 
\eqn\diMEvo{L_1(1,2),\ \ L_2(2,1),\ \ L_3(3,0),\ \ 
\bar\ell_1(2,3),\ \ \bar\ell_2(4,4),\ \ \bar\ell_3(4,2)\ .}
The mass matrices have the form
\eqn\MEvo{M_\nu\sim{\vev{\phi_u}^2\over\Lambda_H}\pmatrix{
\l^6&\l^6&\l^6\cr \l^6&\l^6&\l^6\cr \l^6&\l^6&1\cr},\ \ \ 
M_{\ell^\pm}\sim\vev{\phi_d}
\pmatrix{\l^8&\l^{11}&\l^9\cr \l^8&\l^5&\l^3\cr \l^8&\l^5&\l^3},}
yielding
\eqn\disvo{{\Delta m^2_{1x}\over\Delta m^2_{23}}\sim\l^{12},
\ \ \ s_{12}\sim1,\ \ \ s_{23}\sim1,\ \ \ s_{13}\sim\l^6.}

It is impossible to produce an MSW(LMA) model with $s_{13}\ll1$.
This fact is closely related to the $Z_2$ symmetry which, when
the charges are chosen to give $s_{23}\sim1$ and $s_{12}\sim1$,
always gives $s_{13}\sim1$. More complicated models, {\it e.g.}
models with a $Z_p\times Z_q\times Z_r$ symmetry, can accommodate 
the MSW(LMA) solution as well.

Even though  the examples above all employ an even $m$, 
one can also construct models of mass enhancement with an odd $m$.

\subsec{Mixing Enhancement}
A discrete symmetry can enhance a mixing angle. Take the 
horizontal charges of the second and third generation leptons to be
\eqn\disch{L_2(m-l,l),\ \ \ L_3(0,0),\ \ \ \bar\ell_2(l,5-l),\ \ \ 
\bar\ell_3(l,3-l)\ .}
The mass matrices then have, again, the form \dismas\ and, consequently,
\dispar\ holds.

The mechanism that leads, however, to these results is different
from that of the previous subsection. 
The crucial point here is that 
$(M_{\ell^\pm}^{(2)})_{23}$, which would have been $\O(\l^{m+3})$ under 
a $U(1)$, is enhanced to $\O(\l^3)$ under the discrete symmetry.
The idea, then, is that horizontal charges that would lead to small 
off-diagonal terms with $H=U(1)$, can lead to unsuppressed off-diagonal
terms  with $H=Z_m$. A large hierarchy could occur with either symmetry.

It is, again, straightforward to construct explicit examples for
the MSW(SMA) and the VO solutions, but it is impossible to construct
an MSW(LMA) example with  a suppressed $s_{13}$-mixing.

\subsec{Mass Suppression}
A discrete symmetry can suppress neutrino masses.
Consider models with singlet neutrinos at an intermediate scale
$\Lambda_L$. Entries in $M_N$ could be enhanced, compared to
a $U(1)$ model, by a discrete symmetry. (This mechanism is
the same as the one discussed in section 4.1, except that
now it operates on the singlet neutrinos.) This mass enhancement in the
singlet neutrino sector translates, through the see-saw mechanism,
to mass suppression in the light neutrino sector.

To give a concrete example (for MSW(SMA)), take $m=2$ and the
following charges: 
\eqn\disthree{\eqalign{
L_1(0,2),&\ \ \ L_2(0,0),\ \ \ L_3(0,0),\ \ \ 
\bar\ell_1(0,6),\ \ \ \bar\ell_2(0,5),\ \ \ \bar\ell_3(0,3),\cr
N_1(0,3),&\ \ \ N_2(1,0),\ \ \ N_3(1,0)\ .\cr}}
The mass matrices are of  the form
\eqn\dimath{M_D\sim\vev{\phi_u}\pmatrix{\l^5&\l^3&\l^3\cr
\l^3&\l&\l\cr \l^3&\l&\l\cr},\ \ 
M_N\sim\Lambda_L\pmatrix{\l^6&\l^4&\l^4\cr \l^4&1&1\cr \l^4&1&1\cr},} 
and $M_{\ell^\pm}$ is similar to that of~eq.~\MEmas.
The resulting parameters are the same as in eq.~\dismsw. Note, however,
that the fact that the symmetry is discrete is irrelevant for both
$M_D$ and $M_{\ell^\pm}$. It is only relevant for  $M_N$, and the effect 
is an enhancement of two of its eigenvalues by $\O(\l^{-2})$. This, in turn, 
suppresses two of the light neutrino masses by $\O(\l^2)$,  
thus creating the required hierarchy.

Again, one can use this method to construct models that are
consistent with the VO parameters but not 
(for $H=Z_2\times U(1)$) with the MSW(LMA) parameters.
  
\newsec{Suppressing a Mass with Holomorphic Zeros}

\subsec{An Effective-Two-Generation Mechanism}
We can use holomorphy to strongly suppress a neutrino mass ratio
(compared to the naive estimate). The large mixing angle
arises, in this case, from the charged lepton sector.

For simplicity, we start with a two generation
model to demonstrate the proposed mechanism. The horizontal symmetry is
$U(1)\times U(1)$, with the small breaking parameters both of
$\O(\l)$. The charges are:
\eqn\holcha{L_2(-1,1),\ \ \  L_3(0,0),\ \ \ \bar\ell_2(3,2),
\ \ \ \bar\ell_3(3,0)\ .}
Then, the mass matrices are of  the form
\eqn\holmas{M_\nu^{(2)}\sim{\vev{\phi_u}^2\over\Lambda_H}\pmatrix{
0&0\cr0&1\cr},\ \ \ M_{\ell^\pm}^{(2)}\sim\vev{\phi_d}
\pmatrix{\l^5&\l^3\cr \l^5&\l^3}.}
When we rotate to the charged lepton mass basis, $M_\nu^{(2)}$
assumes the form \Mtwo. We get
\eqn\dispar{m_{\nu_2}=0,\ \ \ m_{\nu_3}\sim{\vev{\phi_u}^2\over\Lambda_H},
\ \ \ s_{23}\sim1\ .}
The zero mass of the light state is a generic feature of this mechanism
in the two generation framework. It is, however, lifted (but remains
suppressed) when the model is extended to include three generations.

The mechanism used here is similar 
in some aspects to the alignment mechanism of 
refs.~\nref\NiSe{Y. Nir and N. Seiberg,
 Phys. Lett. B309 (1993) 337, hep-ph/9304307.}%
\nref\LNSb{M. Leurer, Y. Nir and N. Seiberg,
 Nucl. Phys. B420 (1994) 408, hep-ph/9310320.}%
\refs{\NiSe,\LNSb}. For both it is essential that we
use a $U(1)\times U(1)$ symmetry rather than just a single $U(1)$.
Assume that the symmetry is broken by small parameters of $\O(\l^m,\l^n)$.
Suppose that a certain term carries charges $(H_1,H_2)$ under the
horizontal symmetry. Then, we define the {\it effective} horizontal 
charge of this term, $\hat H$, by
\eqn\Heff{\hat H=mH_1+nH_2 \ .}
The basic idea is that a term with charge $\hat H\geq0$ could still
be forbidden by holomorphy if either of $H_1$ and $H_2$ is negative.

In alignment models, the charges of $Q_1$ and 
$Q_2$ give $\hat H(M_{12})-\hat H(M_{22})=1$, so that naively
$M_{12}/M_{22}\sim\l$. The charges of $\bar u_2$ and $\bar d_2$ are such 
that for $M^u_{12}$ both $H_1$ and $H_2$ are positive and the naive mass ratio 
holds, whereas for $M^d_{12}$ $H_1$ (or $H_2$) is negative 
and the ratio vanishes. In the neutrino models of this
section, $\hat H(L_2)=\hat H(L_3)$. Naively then all entries in $M_\nu^{(2)}$ 
are of the same order of magnitude, but $(H_1,H_2)$ are such that
holomorphy forbids all but $(M_\nu)_{33}$. 

It is amusing to note that holomorphic zeros can lead to both a mixing
angle that is much smaller than the corresponding mass ratio (alignment)
and a mixing angle that is much larger than the corresponding mass ratio
(the neutrino model of this section).

It is not trivial to extend the model to a three generation framework
consistent with \SOL. In particular, we are only able to construct
models that are consistent with large angle solutions to the solar
neutrino problem (MSW(LMA)  and VO in \SOL). In these models, 
$\ne$ and $\nm$
form a pseudo-Dirac neutrino that is lighter than $\nt$. The mass
scale of $\nt$ is then appropriate for \ATM, while the small mass
splitting of the pseudo-Dirac neutrino explains \SOL. The effective
low-energy theory is quite similar to the one described in section~6.2
of ref.~\BHSSW, but the full theory (and, in particular, the mechanism
to create the strong hierarchy) is very different.

To have a viable example of the MSW option, 
take for the first generation charges
\eqn\holcha{L_1(1,0),\ \ \  \bar\ell_1(3,4)\ .}
Then, the mass matrices have the form
\eqn\holmas{M_\nu\sim{\vev{\phi_u}^2\over\Lambda_H}\pmatrix{
\l^2&\l&\l\cr \l&0&0\cr \l&0&1\cr},\ \ \ M_{\ell^\pm}\sim\vev{\phi_d}
\pmatrix{\l^8&\l^6&\l^4\cr \l^7&\l^5&\l^3\cr \l^7&\l^5&\l^3},}
yielding
\eqn\dispar{{\Delta m^2_{1x}\over\Delta m^2_{23}}\sim\l^3,
\ \ \ \sin2\theta_{12}=1-\O(\l^2),\ \ \ s_{23}\sim1,\ \ \ s_{13}\sim\l.}
The $\O(\l^2)$ correction to $\sin2\theta_{12}=1$ should
be quite large to satisfy the upper bound that applies to MSW(LMA),
$\sin2\theta_{12}<0.9$. Note that if the only source for this
correction were $(M_\nu)_{11}/(M_\nu)_{12}=\l$, then it would
be accidentally suppressed, i.e., $\sin2\theta_{12}=1-\l^2/8$.
Therefore, the $\O(\l)$ correction to $s_{12}$ from the
charged lepton mass matrix is important in making this class
of models plausible candidates for the large angle option
of the MSW mechanism.

An example that produces the VO option is the following:
\eqn\holcha{L_1(2,2),\ \ \  \bar\ell_1(6,-2)\ .}
The mass matrices have the form
\eqn\holma{M_\nu\sim{\vev{\phi_u}^2\over\Lambda_H}\pmatrix{
\l^8&\l^4&\l^4\cr \l^4&0&0\cr \l^4&0&1\cr},\ \ \ M_{\ell^\pm}\sim\vev{\phi_d}
\pmatrix{\l^8&\l^9&\l^7\cr 0&\l^5&\l^3\cr 0&\l^5&\l^3},}
yielding
\eqn\dispa{{\Delta m^2_{1x}\over\Delta m^2_{23}}\sim\l^{12},
\ \ \ \sin2\theta_{12}=1-\O(\l^8),\ \ \ s_{23}\sim1,\ \ \ s_{13}\sim\l^4.}

\subsec{A Three Generation Mechanism}
Holomorphic zeros can easily reproduce the interesting three generation
mass matrix \Mthree. In this case one combination of $\nu_\tau$ 
and $\nu_\mu$ is the lightest neutrino while the orthogonal combination
combines with $\nu_e$ to form a pseudo-Dirac neutrino.

Actually, an explicit model is easy to present. Just take the model
of section 3.1 but assume that the full high energy theory has only
singlet neutrinos in vector representations of $H$, namely,  there
are no singlet neutrinos at an intermediate scale (the $\Lambda_L=
\Lambda_H$ case). Then, as explained in section 2, the selection
rules can be applied directly to $M_\nu^{\rm light}$. The 
charges of $L_i$ of eq.~\LNcharges,
\eqn\LNcharges{L_1(y),\ \ L_2(-x),\ \ L_3(-x),\ \ (y\geq x>0)\ ,}
lead to
\eqn\Mligh{M_\nu^{\rm light}\sim{\vev{\phi_u}^2\over\Lambda_H}
\pmatrix{\l^{2y}&\l^{y-x}&\l^{y-x}\cr\l^{y-x}&0&0\cr\l^{y-x}&0&0\cr},}
which is precisely the structure \Mthree. Holomorphy here makes the
whole $2-3$ block of $M_\nu^{\rm light}$ vanish. The matrix
is then of rank 2, so that one neutrino, $-s_{23}\nu_\mu+c_{23}\nu_\tau$
(where $s_{23}=\O(1)$), is rendered massless. For the two components of 
the pseudo-Dirac neutrino, which we denote by $\nu_1$ and $\nu_x$, we have
\eqn\ObsThr{m_{1,x}\sim{\vev{\phi_u}^2\over\Lambda_L}\l^{y-x},\ \ \ 
{\Delta m^2_{1x}\over m^2_{1,x}}\sim\l^{y+x},\ \ \ \sin2\theta_{1x}=
1-\O(\l^{2(y+x)})\ .}
Since the scale of $m_x$ is set by \ATM\ and the mass splitting
$\Delta m^2_{1x}$ is set by \SOL, the ratio between them should be
small and, consequently, so is the deviation of $\sin2\theta_{1x}$
from unity. Therefore, this scenario can only fit the vacuum oscillation
solution of the solar neutrino problem \BHSSW. It requires $y+x\sim10$.

\newsec{Conclusions}
The neutrino flavor parameters that seem to emerge from
the observations of atmospheric and solar neutrinos are
not easily accommodated in flavor models that explain the
smallness and hierarchy in the charged fermion parameters.
In particular, many of these models relate large mixing
in the neutrino sector to non-hierarchical masses, while
the most straightforward explanation of the experimental data
requires $\sin\theta_{23}\sim1$ and $m_{\nu_2}\ll m_{\nu_3}$.

In this work, we investigated the implications of the neutrino
parameters for flavor models with the following features:
(a) Three light neutrinos; (b) Supersymmetry; (c)
Abelian horizontal symmetry broken by a single parameter.

Within this framework, several mechanisms for obtaining
large mixing together with a large hierarchy were suggested:
\item{(i)} Accidental hierarchy.
\item{(ii)} $R_p$ violation: different neutrino generations
 acquire their masses from different sources.
\item{(iii)} See-saw mass enhancement: a light neutrino mass is 
 enhanced because it is acquired through a see-saw mechanism, and 
 the corresponding singlet neutrino mass is suppressed. 

Here we suggested several new mechanisms:
\item{(iv)} Mass enhancement from discrete symmetries: a light neutrino
 mass could be larger for a $Z_n$ symmetry than its would-be value
 for a $U(1)$ symmetry.
\item{(v)} Mixing enhancement from discrete symmetries: a mixing angle
 could be larger for a $Z_n$ symmetry than its would-be value
 for a $U(1)$.
\item{(vi)} Mass suppression from discrete symmetries: a singlet-neutrino 
 mass is enhanced by a discrete symmetry, thus suppressing a light 
 neutrino mass through the see-saw mechanism.
\item{(vii)} Holomorphic zeros: a neutrino mass is suppressed by
 holomorphic zeros in the mass matrix.

Outside our framework of supersymmetric Abelian horizontal symmetries
there are, of course, many other proposals in the literature (some
related to various symmetries and some which are simply ansatze for
mass matrices) for  achieving a large hierarchy together with large 
mixing~\nref\DLLRS{H. Dreiner, G.K. Leontaris, 
S. Lola, G.G. Ross and C. Scheich,
Nucl.Phys. B436 (1995) 461, hep-ph/9409369.}%
\nref\LLR{G.K. Leontaris, S. Lola and G.G. Ross,
 Nucl.Phys. B454 (1995) 25, hep-ph/9505402.}%
\nref\BeRo{Z. Berezhiani and A. Rossi,
 Phys. Lett. B367 (1996) 219, hep-ph/9507393.}%
\nref\LLSV{G.K. Leontaris, S. Lola, C. Scheich and J. Vergados,
 Phys.Rev. D53 (1996) 6381, hep-ph/9509351.}%
\nref\BPWW{V. Barger, S. Pakvasa, T.J. Weiler and K. Whisnant,
 hep-ph/9806387.}%
\nref\BGG{A.J. Baltz, A.S. Goldhaber and M. Goldhaber, hep-ph/9806540.}%
\nref\FrXi{H. Fritzsch and X. Xing, hep-ph/9808272.}%
\nref\Eyal{G. Eyal, hep-ph/9807308.}%
\nref\Jez{M. Jezabek and Y. Sumino, hep-ph/9807310.}%
\nref\AlFe{G. Altarelli and F. Feruglio, hep-ph/9807353.}%
\nref\eMa{E. Ma, hep-ph/9807386.}%
\nref\Tani{M. Tanimoto, hep-ph/9807517.}%
\nref\Haba{N. Haba, hep-ph/9807552.}%
\nref\JoVe{A.S. Joshipura and S.K. Vempati, hep-ph/9808232.}%
\nref\OTTY{K. Oda, E. Takasugi, M. Tanaka and M. Yoshimura,
 hep-ph/9808241.}%
\nref\GeGl{H. Georgi and S.L. Glashow, hep-ph/9808293.}%
\nref\MoNu{R.N. Mohapatra and S. Nussinov, hep-ph/9808301.}%
\nref\BHS{R. Barbieri, L.J. Hall and A. Strumia, hep-ph/9808333.}%
\refs{\DLLRS-\BHS}.

\bigskip
\noindent
{\bf Acknowledgements}
\smallskip
\noindent
We thank Lance Dixon and Nati Seiberg for helpful discussions.
Y.N. and Y.S. are supported in part by the
United States -- Israel Binational Science Foundation (BSF).
Y.N. is supported in part by the Israel Science
Foundation and by the Minerva Foundation (Munich).

\listrefs
\end